\documentclass[aps,prx,reprint,superscriptaddress,longbibliography]{revtex4-2}
\usepackage{amssymb} 
\usepackage{amsmath}
\usepackage{graphicx}
\usepackage{fancyhdr}
\usepackage{ifthen}
\usepackage{multirow}
\usepackage{footmisc}
\usepackage{array}
\usepackage{color}
\usepackage{ragged2e}
\usepackage{CJKutf8}
\usepackage{hyperref}
\usepackage{physics}

\usepackage{lipsum}
\DeclareSymbolFont{myletters}{OML}{ztmcm}{m}{it}
\DeclareMathSymbol{\uplambda}{\mathord}{myletters}{"15}

\usepackage[]{hyperref}
\usepackage{xcolor}
\definecolor{LinkColor}{RGB}{195,57,57}
\hypersetup{colorlinks=true,citecolor=LinkColor,linkcolor=LinkColor,urlcolor=LinkColor}

\newcounter{aq}
\definecolor{darkgreen}{RGB}{0, 128, 0}

{}

\begin{document}
\title{Unique Hierarchical Rotational Dynamics Induces Ultralow Lattice Thermal Conductivity in Cyanide-bridged Framework Materials}
\author{Zhunyun Tang}
\date{\today}
\affiliation{School of Physics and Optoelectronics $\&$ Hunan Provincial Key Laboratory of Computational Condensed Matter Physics and Quantum Materials Engineering, Xiangtan University, Xiangtan 411105, Hunan, China}

\author{Xiaoxia Wang }
\affiliation{School of Physics and Optoelectronics $\&$ Hunan Provincial Key Laboratory of Computational Condensed Matter Physics and Quantum Materials Engineering, Xiangtan University, Xiangtan 411105, Hunan, China}

\author{Jin Li}
\affiliation{School of Physics and Optoelectronics $\&$ Hunan Provincial Key Laboratory of Computational Condensed Matter Physics and Quantum Materials Engineering, Xiangtan University, Xiangtan 411105, Hunan, China}

\author{Chaoyu He}
\affiliation{School of Physics and Optoelectronics $\&$ Hunan Provincial Key Laboratory of Computational Condensed Matter Physics and Quantum Materials Engineering, Xiangtan University, Xiangtan 411105, Hunan, China}

\author{Mingxing Chen}
\affiliation{School of Physics and Electronics, Hunan Normal University, Changsha 410081, Hunan, China}

\author{Chao Tang}
\thanks{\textnormal{} 
\href{mailto:tang\_chao@xtu.edu.cn}{tang\_chao@xtu.edu.cn}.}
\affiliation{School of Physics and Optoelectronics $\&$ Hunan Provincial Key Laboratory of Computational Condensed Matter Physics and Quantum Materials Engineering, Xiangtan University, Xiangtan 411105, Hunan, China}

\author{Tao Ouyang}
\thanks{\textnormal{} 
\href{mailto:ouyangtao@xtu.edu.cn}{ouyangtao@xtu.edu.cn}.}
\affiliation{School of Physics and Optoelectronics $\&$ Hunan Provincial Key Laboratory of Computational Condensed Matter Physics and Quantum Materials Engineering, Xiangtan University, Xiangtan 411105, Hunan, China}

\begin{abstract}
The pursuit of materials combining light constituent elements with ultralow lattice thermal conductivity ($\kappa_{\mathrm{L}}$) is crucial to advancing technologies like thermoelectrics and thermal barrier coatings, yet it remains a formidable challenge to date. Herein, we achieve ultralow $\kappa_{\mathrm{L}}$ in lightweight cyanide-bridged framework materials (CFMs) through the rational integration of properties such as the hierarchical vibrations exhibited in superatomic structures and rotational dynamics exhibited in perovskites. Unique hierarchical rotation behavior leads to multiple negative peaks in Grüneisen parameters across a wide frequency range, thereby inducing pronounced negative thermal expansion and strong cubic anharmonicity in CFMs. Meanwhile, the synergistic effect between large four-phonon scattering phase space (induced by phonon quasi-flat bands and wide bandgaps) and strong quartic anharmonicity (associated with rotation modes) leads to giant quartic anharmonic scattering rates in these materials. Consequently, the $\kappa_{\mathrm{L}}$ of these CFMs decreases by one to two orders of magnitude compared to the known perovskites or perovskite-like materials with equivalent average atomic masses. For instance, the Cd(CN)$_{2}$, NaB(CN)$_{4}$, LiIn(CN)$_{4}$, and AgX(CN)$_{4}$ (X = B, Al, Ga, In) exhibit ultralow room-temperature $\kappa_{\mathrm{L}}$ values ranging from 0.35 to 0.81 W/mK. This work not only establishes CFMs as a novel and rich platform for studying extreme phonon anharmonicity, but also provides a new paradigm for achieving ultralow thermal conductivity in lightweight materials via the conscious integration of hierarchical and rotational dynamics. 
\end{abstract}

\maketitle



\section{Introduction}\label{sec:1}	
The pursuit of materials with ultralow lattice thermal conductivity ($\kappa_{\mathrm{L}}$) is a central endeavor in condensed matter physics and materials science, driven by applications in thermoelectrics, thermal insulation, and barrier coatings \cite{RN1,RN2,RN3,RN4}. Conventional strategies for suppressing $\kappa_{\mathrm{L}}$ often rely on the incorporation of heavy constituent elements or structural complexity \cite{RN5,RN6,RN7,RN8}, which typically conflicts with the demand for lightweight functional materials in applications ranging from aerospace components to flexible electronics. Consequently, the exploration of lightweight materials that simultaneously exhibit ultralow $\kappa_{\mathrm{L}}$ remains a formidable goal, one that requires moving beyond conventional phonon scattering paradigms toward designs that leverage emergent lattice dynamics and anharmonicity.

Recent advances highlight the role of specific bonding characteristics and dynamic structural features in driving phonon anharmonicity and collapse of thermal transport \cite{RN9,RN10,RN11,RN12,RN13,RN14,RN15,RN16}. Specifically, the presence of anti-bonding valence states leads to bond weakening and formation of soft optical phonons, resulting in intrinsic phonon scattering rates far exceeding those in conventional semiconductors \cite{RN9,RN10}. Alternatively, the concept of hierarchical vibrational architectures (inspired by superatomic crystals) has emerged as a potent design strategy \cite{RN11}. Hierarchical vibrations endow additional rotational or translational degrees of freedom, leading to strong coupling between internal optical phonons and acoustic modes, further generating extreme phonon anharmonic scattering \cite{RN12,RN13,RN14,RN15,RN16}. In a parallel development, it has been found that low-energy optical modes associated with octahedral rotations in perovskites (perovskite-like materials) often exhibit strong anharmonicity, leading to low $\kappa_{\mathrm{L}}$ and pronounced wave-like tunneling behavior \cite{RN17,RN18,RN19,RN20,RN21}.

Although these mechanisms have been individually established in disparate material classes, a fundamental question remains: can these distinct phonon-suppressing features be intentionally combined within a lightweight material system to achieve ultralow thermal conductivity through synergistic effects? Based on this design principle, we perform a systematical search in the Inorganic Crystal Structure Database (ICSD) and discover that cyanide-bridged framework materials (CFMs) present an exceptionally promising platform for such an integrated approach \cite{RN22}. Their framework structures, comprising M---C$\equiv$N---M$'$ linkages, the flexibility of the cyanide bridge allows for dynamic disorder and local distortions, reminiscent of perovskites, while the subunits within the framework suggests a pathway toward hierarchical vibration \cite{RN23,RN24,RN25}. Despite these intriguing features, their thermal transport properties remain largely unexplored, particularly in relation to the interplay among local rotational degrees of freedom, hierarchical vibrations, and collective low-energy dynamics. 

In this study, we fill this gap by employing first-principles calculations combined with machine learning potentials. To ensure reliability, the thermal conductivity values are cross-validated through unified phonon theory and large-scale molecular dynamics simulations. The results demonstrate that this family of materials uniquely combines the key phonon-suppressing mechanisms previously identified in separate systems: the hierarchical vibrational architecture characteristic of superatomic crystals and the rotational dynamics typical of perovskites. This cooperative mechanism leads to emergent phenomena, including multiple negative Grüneisen parameter peaks across a broad frequency range, signaling pronounced negative thermal expansion and strong cubic anharmonicity. Most notably, we identify a synergistic interplay between the large four-phonon scattering phase space (arising from phonon quasi-flat bands and wide bandgaps) and the strong quartic anharmonicity (associated with rotation modes), leading to giant four-phonon scattering rates that dominate thermal resistance. As a result, the $\kappa_{\mathrm{L}}$ of these CFMs is suppressed by one to two orders of magnitude compared to the previously known perovskite and perovskite-like materials with equivalent average atomic mass. Specific compounds, such as Cd(CN)$_{2}$, NaB(CN)$_{4}$, LiIn(CN)$_{4}$, and AgX(CN)$_{4}$ (X = B, Al, Ga, In), exhibit ultralow room-temperature $\kappa_{\mathrm{L}}$ values ranging from 0.35 to 0.81 W/mK, positioning them among the lowest known thermal conductivity materials for their weight class. These findings provide a promising strategy for designing lightweight materials with ultralow thermal conductivity.

\section{Results and Discussion}\label{sec:2}
\begin{figure*}[t!] 
\centering
	\includegraphics[width=1.0\linewidth]{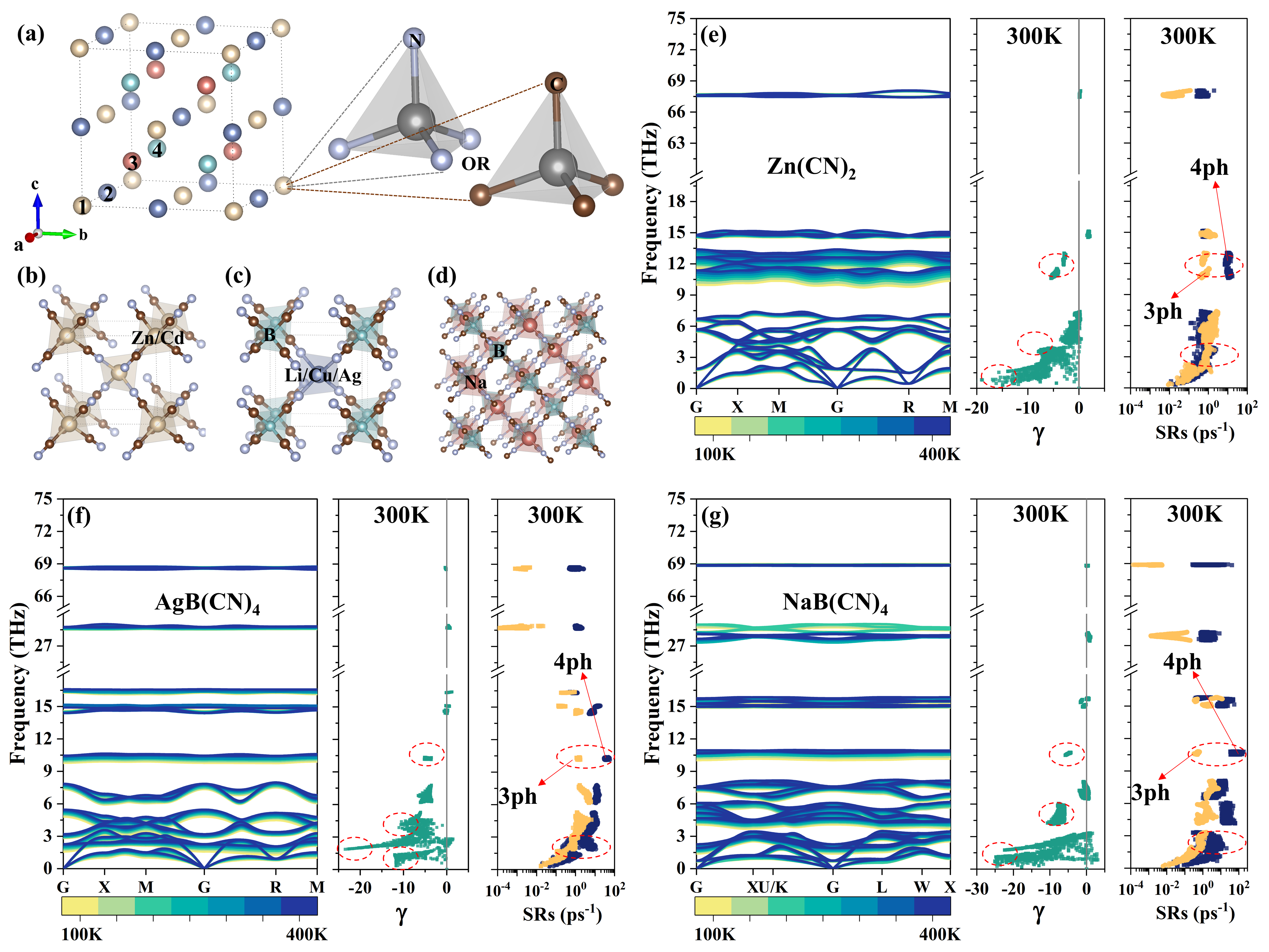}
	\caption{(a) A simplified model of cubic crystal structure, with each sphere consisting of a tetrahedral superatom cluster. Scheme of unit cell structure of (b) Zn(CN)$_{2}$ and Cd(CN)$_{2}$, (c) LiB(CN)$_{4}$, CuB(CN)$_{4}$, and AgB(CN)$_{4}$, (d) and NaB(CN)$_{4}$. (e-g) Calculated finite-temperature phonon dispersion, accounting for thermal expansion effects, for Zn(CN)$_{2}$, AgB(CN)$_{4}$, and NaB(CN)$_{4}$, respectively. The middle and right panels of (e-g) present the mode $\gamma$ and SRs (including three- and four-phonon processes) at room temperature, respectively.}
	\label{fgr:fig-1}
\end{figure*}

\begin{figure*}[t!] 
	\centering
	\includegraphics[width=1.0\linewidth]{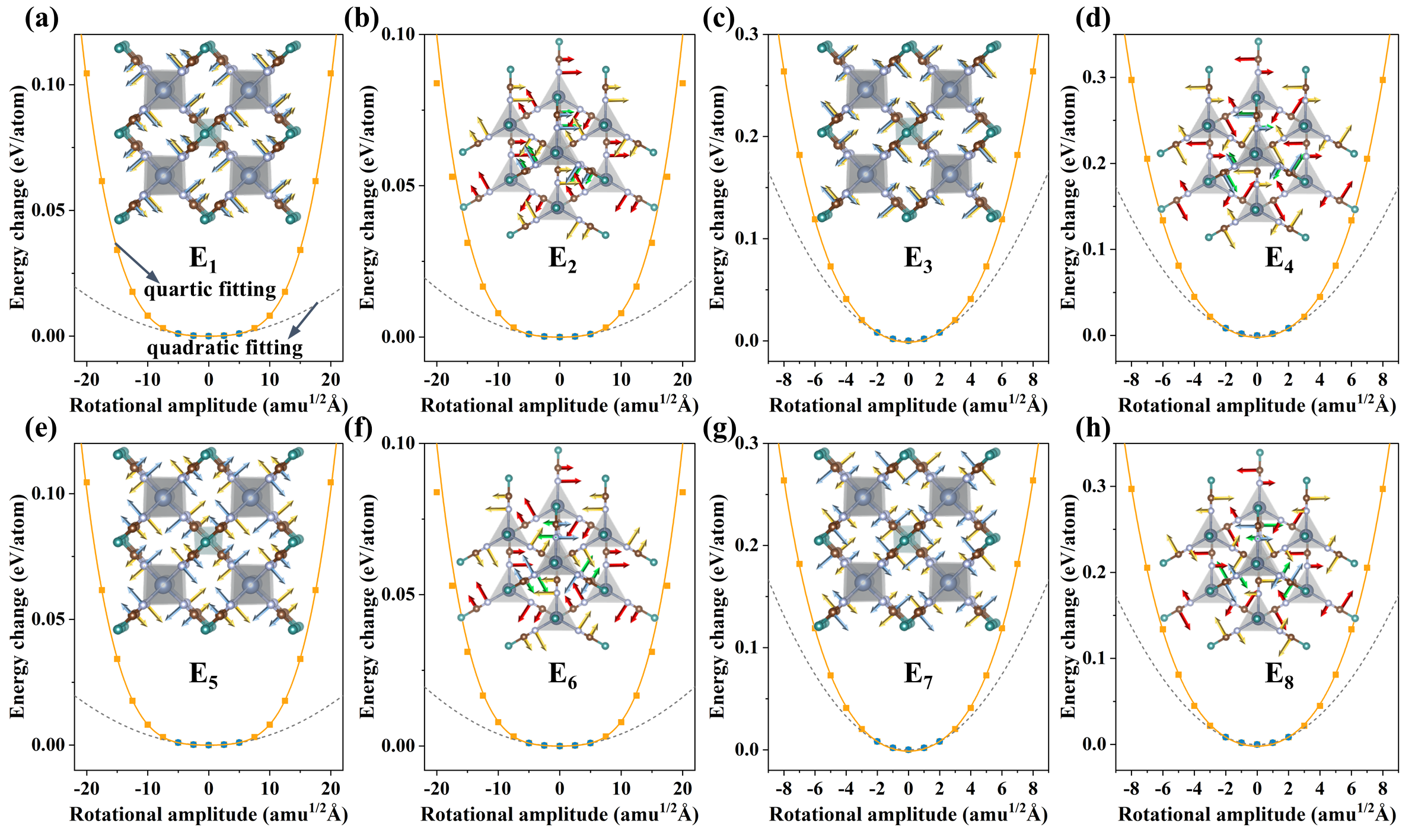}
	\caption{The potential well for different rotation modes. The quadratic and quartic fitting curves are represented by gray dashed lines and yellow solid lines, respectively. The insets label the vibration eigenvectors for different rotation modes}
	\label{fgr:fig-2}
\end{figure*}
\begin{figure}[t!] 
	\centering
	\includegraphics[width=1.0\linewidth]{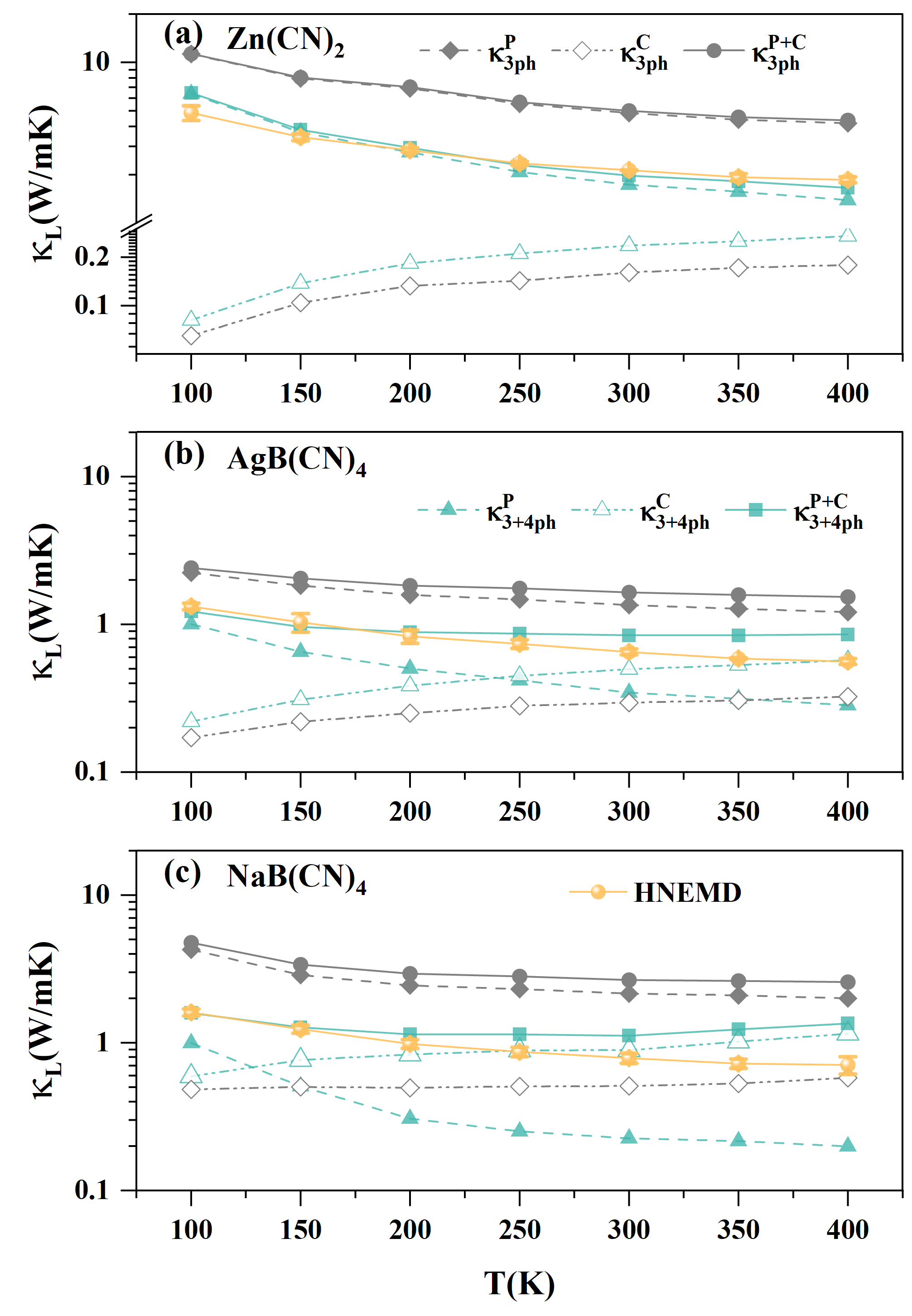}
	\caption{The temperature-dependent $\kappa_{\mathrm{L}}$ of (a) Zn(CN)$_{2}$, (b) AgB(CN)$_{4}$, and (c) NaB(CN)$_{4}$ is calculated based on different models are compared with molecular dynamics simulations.}
	\label{fgr:fig-3}
\end{figure}

Figure~\ref{fgr:fig-1}(a) illustrates a simplified model of cubic-phase atomic structure, where four distinct Wyckoff positions are represented by differently colored spheres. When positions 1 and 2 are occupied by specific tetrahedral superatomic clusters ZnC$_{4}$ or CdC$_{4}$, and positions 3 and 4 by ZnN$_{4}$ or CdN$_{4}$ clusters, these units connect through ---C$\equiv$N--- bridges to form the classic Zn(CN)$_{2}$ or Cd(CN)$_{2}$ framework materials with P$-$43m space group, as depicted in Fig.~\ref{fgr:fig-1}(b). If positions 1 and 2 are occupied by BC$_{4}$ clusters and positions 3 and 4 by AgN$_{4}$, CuN$_{4}$, or LiN$_{4}$ clusters, the resulting structure corresponds to AgB(CN)$_{4}$, CuB(CN)$_{4}$, or LiB(CN)$_{4}$, respectively, all crystallizing in the P$-$43m space group, as shown in Fig.~\ref{fgr:fig-1}(c). In contrast, when positions 1 and 4 are occupied by BC$_{4}$ clusters and positions 2 and 3 by NaN$_{4}$ clusters, the framework adopts the Fd$-$3m space group, characteristic of NaB(CN)$_{4}$, see Fig.~\ref{fgr:fig-1}(d). Previous experimental and theoretical studies have extensively explored the thermal expansion properties of these cyanide-bridged framework materials (CFMs) \cite{RN25,RN26,RN27,RN28,RN29,RN30}.  The results reveal that these compounds generally exhibit negative thermal expansion (NTE) behavior. Figure S2 of the Supporting Information (SI) demonstrates that the calculated NTE behavior exhibits markedly improved agreement with experimental data when the van der Waals (vdW) correction term is included in molecular dynamics (MD) simulations. This indicates that intercluster vdW interactions are crucial for accurately capturing the lattice dynamics and anharmonicity that drive the pronounced NTE in these cyanide frameworks. Notably, despite their low average atomic mass, Zn(CN)$_{2}$, Cd(CN)$_{2}$, AgB(CN)$_{4}$, and NaB(CN)$_{4}$ hold exceptionally pronounced NTE effects, surpassing most other known NTE materials \cite{RN31}. Such remarkable behavior foreshows the presence of intriguing lattice dynamics in these CFMs. Therefore, we select Zn(CN)$_{2}$, AgB(CN)$_{4}$, and NaB(CN)$_{4}$ as representative systems and further computed their phonon spectra. As illustrated in Figs.~\ref{fgr:fig-1}(e-g), the presence of light elements leads to remarkably high optical phonon cutoff frequencies. The hierarchical vibrational behavior inherent to superatomic structures gives rise to a plethora of localized phonon modes, inducing numerous phonon quasi-flat bands and wide bandgaps. Moreover, certain phonon modes exhibit pronounced temperature-dependent behavior, particularly the optical branches around 2–6 THz and 10 THz, suggesting strong anharmonic effects in these specific modes. In conventional perovskite or perovskite-like NTE materials, great NTE coefficients mainly originate from large negative Grüneisen parameters associated with extremely low-frequency soft modes linked to octahedral rotations \cite{RN21,RN32,RN33}. However, these traditional NTE materials typically exhibit only a limited set of rotation phonon modes due to constrained rotational degrees of freedom. In contrast, the unique hierarchical vibrations confer additional rotational freedom in CFMs, resulting in richer rotation phonon modes, as depicted in the insets of Fig.~\ref{fgr:fig-2} (and Fig. S7 of the SI). The positions of these rotation modes are marked on the phonon dispersion in Fig. S8 of the SI, with AgB(CN)$_{4}$ as an example. The distinctive hierarchical rotation behavior leads to multiple negative peaks in the Grüneisen parameter ($\gamma$) across a broad frequency range, as shown in the middle panels of Figs.~\ref{fgr:fig-1}(e-g). This widespread negative $\gamma$ distribution significantly enhances the overall cubic anharmonicity and is a key driver of the pronounced NTE observed in these CFMs. Moreover, by mapping the potential energy curves along the coordinates of these rotation modes, one can find that the actual potential deviates significantly from the harmonic approximation (quadratic fitting), as demonstrated in Figs.~\ref{fgr:fig-2}(a–h). It is worth noting that the fitted curves agree well with the DFT data points when the quartic term is introduced. This deviation provides direct evidence of strong quartic anharmonicity associated with these rotation modes. Additionally, the phonon quasi-flat bands and wide bandgaps prevalent in these materials generally expand the available phase space for four-phonon scattering processes \cite{RN34,RN35}, as shown in Fig. S9 of the SI. As a result, the synergistic effect between large four-phonon scattering phase space and intrinsically strong quartic anharmonicity leads to giant four-phonon anharmonic scattering rates (SRs) in these materials, as quantitatively shown in the right panels of Figs.~\ref{fgr:fig-1}(e-g). 

\begin{figure}[t!] 
	\centering
	\includegraphics[width=1.0\linewidth]{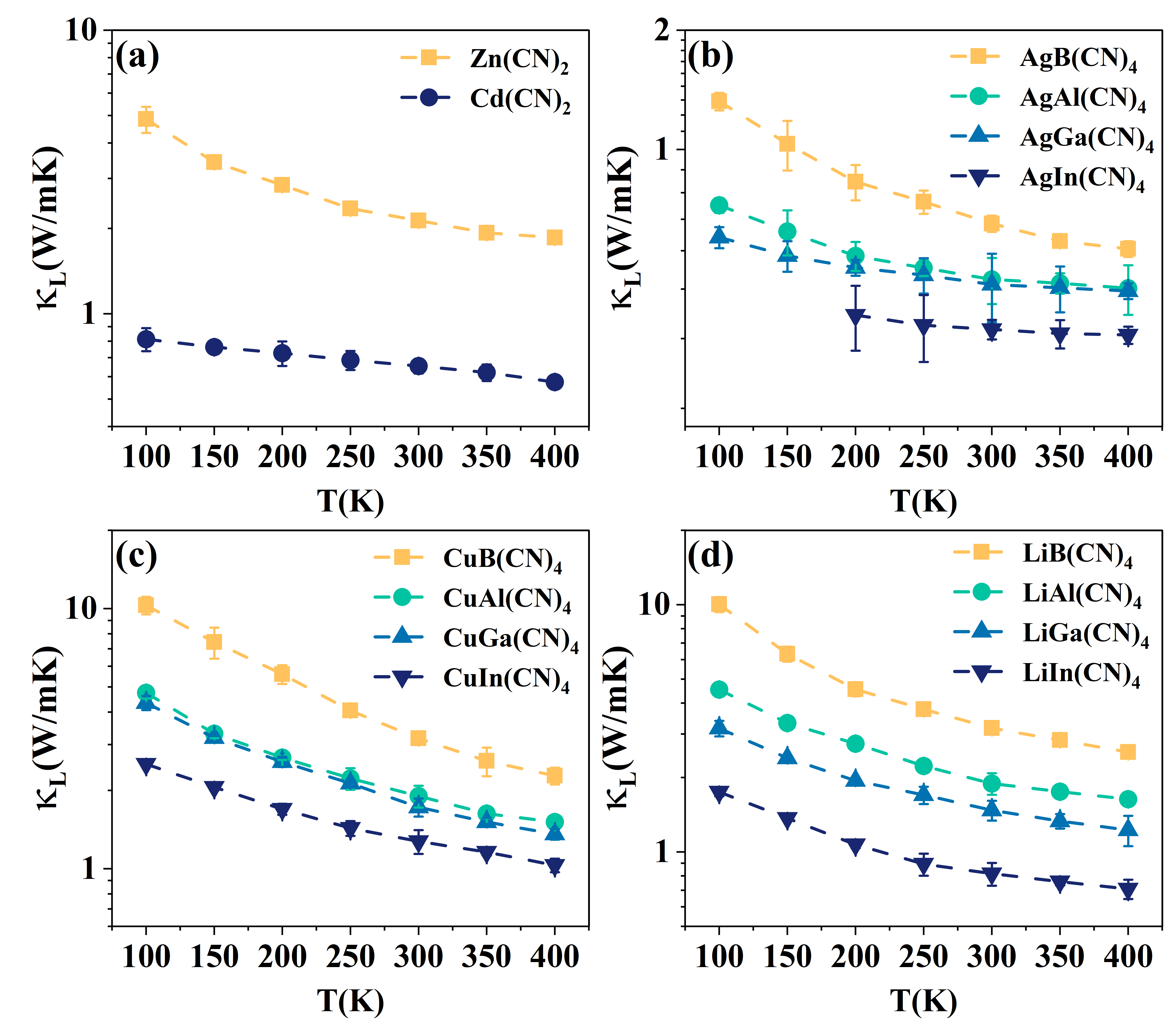}
	\caption{The temperature-dependent $\kappa_{\mathrm{L}}$ calculated using the HNEMD method for (a) Zn(CN)$_{2}$ and Cd(CN)$_{2}$, (b) AgX(CN)$_{4}$, (c) CuX(CN)$_{4}$, and (d) LiX(CN)$_{4}$ (X=B, Al, Ga, In).}
	\label{fgr:fig-4}
\end{figure}
\begin{figure}[t!] 
	\centering
	\includegraphics[width=1.0\linewidth]{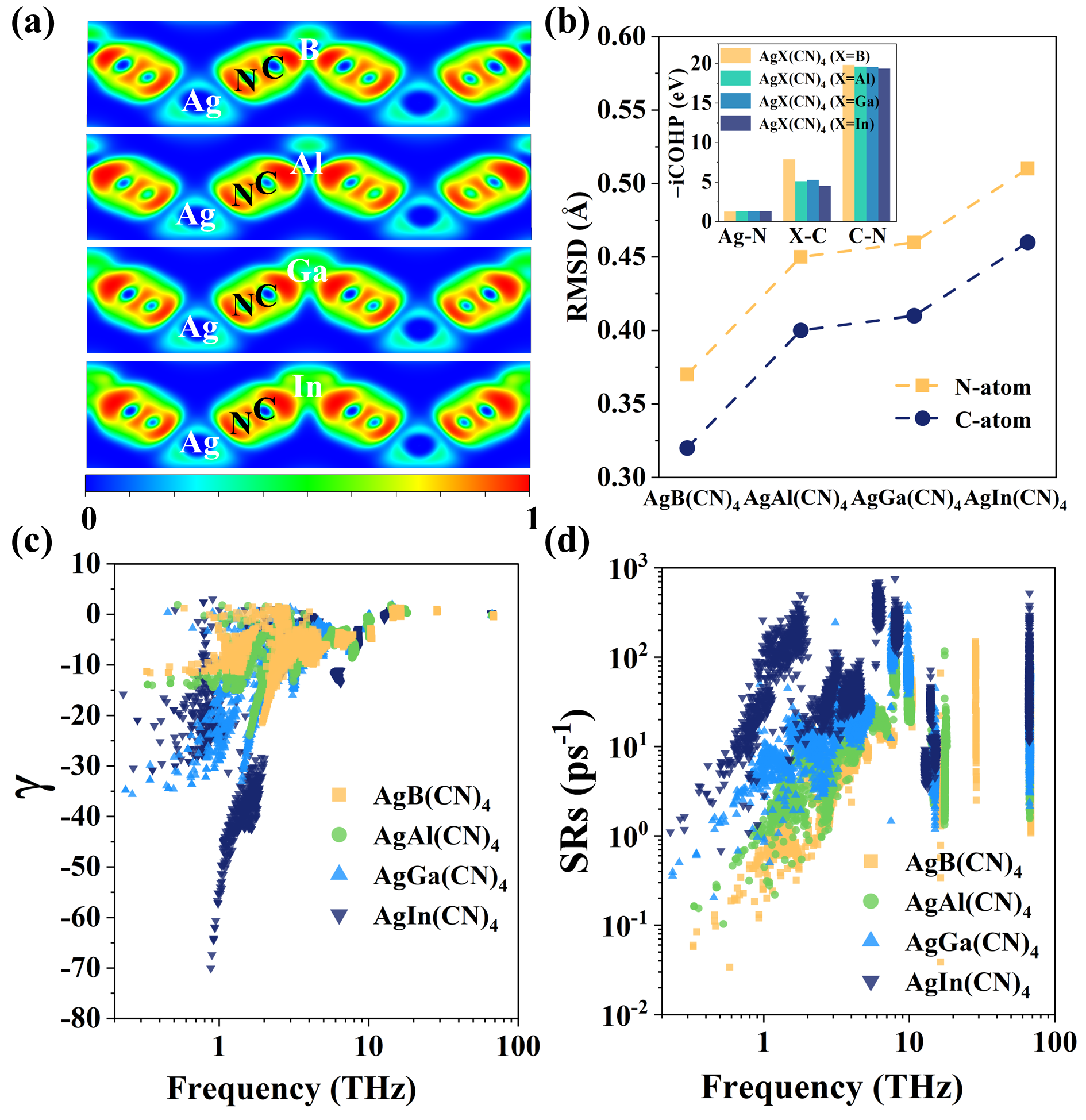}
	\caption{(a) The ELF, (b) RMSD of N and C atoms, (c) $\gamma$, and (d) phonon SRs for AgX(CN)$_{4}$ (X=B, Al, Ga, In). The inset of (b) shows the –iCOHP for different atomic bonds.}
	\label{fgr:fig-5}
\end{figure}
\begin{figure}[t!] 
	\centering
	\includegraphics[width=1.0\linewidth]{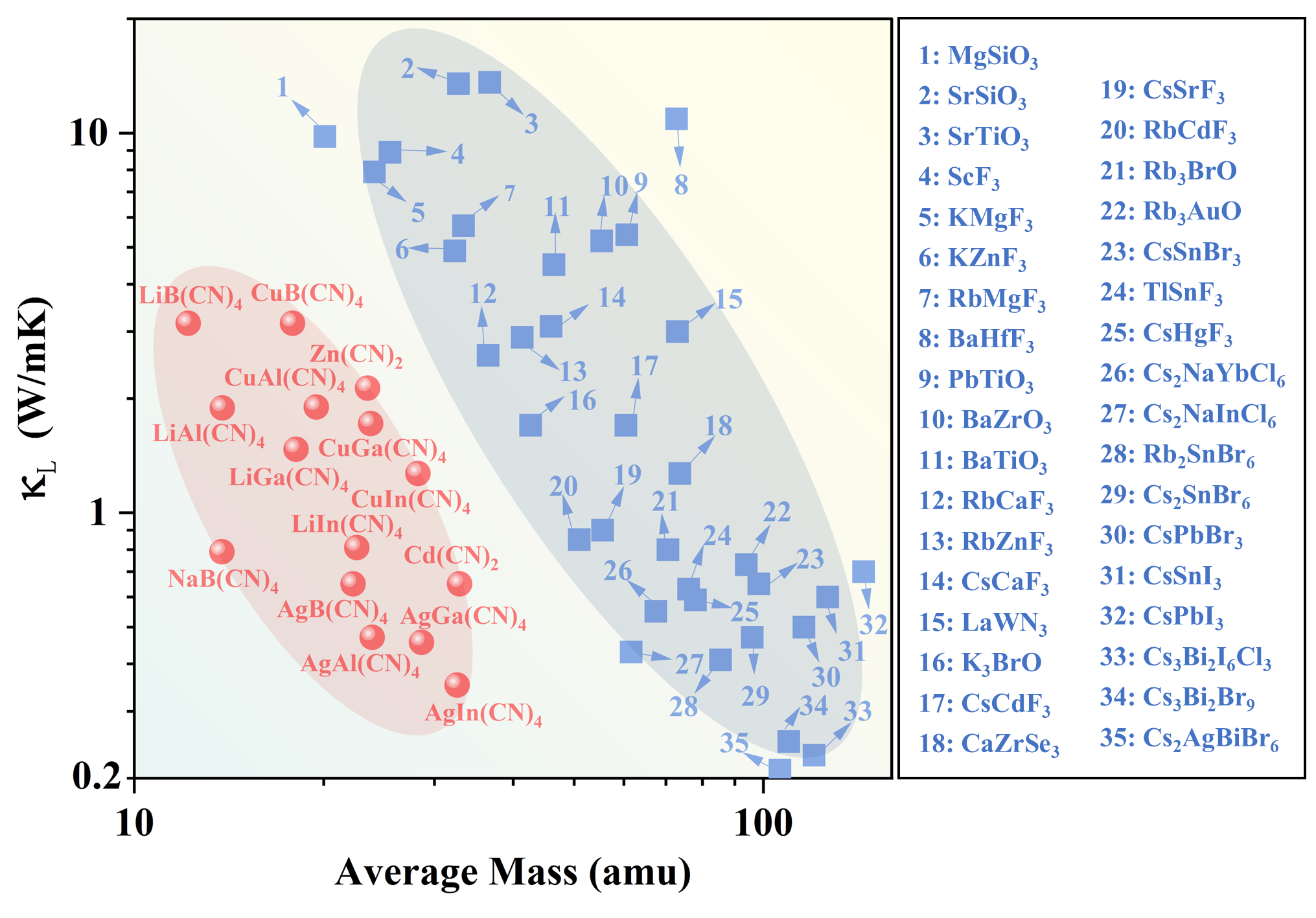}
	\caption{The dependence of room temperature $\kappa_{\mathrm{L}}$ on average mass. The red spheres mark the 15 CFMs studied in this paper. The blue squares mark some typical perovskite and perovskite-like materials.}
	\label{fgr:fig-6}
\end{figure}

We now turn our attention to the thermal transport properties of these CFMs. First, the lattice thermal conductivity ($\kappa_{\mathrm{L}}$) of Zn(CN)$_{2}$, AgB(CN)$_{4}$, and NaB(CN)$_{4}$ in the temperature range of 100 to 400 K is obtained based on the self-consistent phonon (SCP) approach combined with the unified phonon transport theory \cite{RN36,RN37}, accounting of the thermal expansion effect. The results are decomposed into particle-like ($\kappa_{\mathrm{L}}^{\mathrm{P}}$) and wave-like coherence ($\kappa_{\mathrm{L}}^{\mathrm{C}}$) contributions, with and without four-phonon (4ph) scattering. All three materials exhibit low $\kappa_{\mathrm{L}}$ values, which is consistent with the strong anharmonicity and enhanced scattering discussed above. In Zn(CN)$_{2}$, the inclusion of four-phonon processes drastically reduces $\kappa_{\mathrm{L}}^{\mathrm{P}}$ from 4.84 (3ph only; $\kappa_{\mathrm{3ph}}^{\mathrm{P}}$) to 1.74 W/mK (3+4ph; $\kappa_{\mathrm{3+4ph}}^{\mathrm{P}}$) at 300 K, with a 64\% suppression [Fig.~\ref{fgr:fig-3}(a)]. Such suppression is even more pronounced in AgB(CN)$_{4}$ and NaB(CN)$_{4}$, where $\kappa_{\mathrm{L}}^{\mathrm{P}}$ is reduced by 75\% (from 1.35 to 0.35 W/mK) and 90\% (from 2.15 to 0.22 W/mK), respectively [Figs.~\ref{fgr:fig-3}(b-c)]. The coherence contribution $\kappa_{\mathrm{3+4ph}}^{\mathrm{C}}$ remains minor in Zn(CN)$_{2}$, accounting for only ~10\% of the total $\kappa_{\mathrm{3+4ph}}^{\mathrm{P+C}}$ at 300 K. However, in AgB(CN)$_{4}$ and NaB(CN)$_{4}$, $\kappa_{\mathrm{3+4ph}}^{\mathrm{C}}$ contributes significantly to the total thermal conductivity, even exceeding $\kappa_{\mathrm{3+4ph}}^{\mathrm{P}}$ at elevated temperatures. Consequently, $\kappa_{\mathrm{L}}$ in the two systems is increasingly dominated by wave-like phonon tunneling, leading to a weak temperature dependence. Remarkably, NaB(CN)$_{4}$ exhibits a positive temperature dependence of $\kappa_{\mathrm{L}}$ at high temperatures, a behavior previously observed in other ultralow $\kappa_{\mathrm{L}}$ materials \cite{RN38,RN39,RN40,RN41}. This is mainly because the stronger anharmonic scattering and smaller average phonon band spacing in NaB(CN)$_{4}$ induce more phonon modes with lifetimes below the Wigner limit \cite{RN42} (see Fig. S10 of the SI), resulting in glass-like thermal transport. However, we also observe that certain extreme phonon modes of NaB(CN)$_{4}$ enter the overdamped regime at high temperatures (Fig. S10 of the SI), which might render the standard phonon gas model invalid \cite{RN43,RN44}. Therefore, we further employ homogeneous nonequilibrium molecular dynamics (HNEMD) simulations to provide a more robust description for thermal transport \cite{RN45}. The HNEMD results are in good agreement with the unified transport theory calculations at low temperatures, but reveal a continuing decrease in $\kappa_{\mathrm{L}}$ at high temperatures, underscoring the limitations of the standard phonon gas model in giant anharmonic systems. Finally, we extend HNEMD simulations to 15 CFMs, including Zn(CN)$_{2}$, Cd(CN)$_{2}$, NaB(CN)$_{4}$, LiX(CN)$_{4}$, CuX(CN)$_{4}$, and AgX(CN)$_{4}$ (X=B, Al, Ga, In). As summarized in Fig.~\ref{fgr:fig-4}, thermal transport is further suppressed as the X atom transitions from B to In. Several compounds [Cd(CN)$_{2}$, NaB(CN)$_{4}$, LiIn(CN)$_{4}$, and all AgX(CN)$_{4}$ members] exhibit ultralow room-temperature values between 0.35 and 0.81 W/mK. Notably, NaX(CN)$_{4}$ (X=Al, Ga, In) are excluded due to their complex thermal expansion behavior (possibly involving multiple structural phase transitions), as shown in Fig. S3 of the SI, which is beyond the scope of this study. In contrast, AgIn(CN)$_{4}$ shows monotonic positive thermal expansion at low temperatures, but transitions to negative thermal expansion above 200 K. This appears to arise from the low-temperature instability of the cubic phase, as evidenced by imaginary phonon modes below 200 K that vanish at higher temperatures [Fig. S11 of the SI], mirroring behavior observed in other cubic frameworks \cite{RN18,RN46}.

To elucidate the microscopic mechanism underlying the gradual suppression of thermal conductivity as the X atom transitions from B to In in these CFMs, we analyze the evolution of chemical bonding and its impact on lattice anharmonicity. Figure~\ref{fgr:fig-5}(a) presents the electron localization function (ELF) for AgX(CN)$_{4}$ (X=B, Al, Ga, In). The intense localization between B and C atoms in AgB(CN)$_{4}$ signifies strong bonding character of the B–C bond. However, as B transitions to In, this localization gradually diminishes, suggesting that the bonding strength of X–C is progressively weakened. To further quantify the bonding strength, the integrated crystal orbital Hamilton population (iCOHP) is calculated \cite{RN47}, a measure of bond strength, for key atomic pairs, as shown in inset of Fig.~\ref{fgr:fig-5}(b). The –iCOHP values for the Ag–N and C–N bonds remain essentially constant throughout the series. However, the –iCOHP for the X–C bond decreases substantially from B to In, confirming a systematic chemical bond weakening. As previously established, the strong anharmonicity in these materials is intimately linked to rotation modes, which primarily involve large-amplitude transverse vibrations of the N and C atoms. The root-mean-square displacements (RMSD) of C and N atoms, calculated at 300 K, are presented in Fig.~\ref{fgr:fig-5}(b). The RMSD values are significantly enhanced in AgIn(CN)$_{4}$ compared to AgB(CN)$_{4}$, indicating more pronounced vibrational amplitudes and thus larger dynamic tetrahedral distortion. This enhanced flexibility directly leads to stronger lattice anharmonicity, which is quantitatively confirmed by the markedly increased negative Grüneisen parameters in AgIn(CN)$_{4}$ [see Fig.~\ref{fgr:fig-5}(c)]. Consequently, AgIn(CN)$_{4}$ exhibits more pronounced negative thermal expansion (Fig. S3 of the SI) and stronger phonon scattering rates (SRs) [Fig.~\ref{fgr:fig-5}(d)], thereby further suppression of thermal transport properties. 

Finally, we plot the room-temperature thermal conductivity ($\kappa_{\mathrm{L}}$) of these CFMs as a function of average atomic mass ($\bar{M}$) in Fig.~\ref{fgr:fig-6}, alongside data for classic perovskites and perovskite-like materials for comparison \cite{RN17,RN19,RN20,RN21,RN38,RN48,RN49,RN50,RN51,RN52,RN53,RN54,RN55,RN56,RN57,RN58,RN59,RN60,RN61,RN62,RN63,RN64}. Obviously, the $\bar{M}$–$\kappa_{\mathrm{L}}$ trend of CFMs deviates significantly from that of perovskite materials. Despite possessing relatively low $\bar{M}$, the CFMs still exhibit ultralow $\kappa_{\mathrm{L}}$, roughly 1-2 orders of magnitude lower than those of perovskites with comparable $\bar{M}$.

\section{Conclusion}\label{sec:3}
In summary, we have unveiled the microscopic origin of ultralow thermal conductivity ($\kappa_{\mathrm{L}}$) in lightweight cyanide-bridged framework materials (CFMs), challenging the conventional trade-off between low average atomic mass and efficient thermal transport. Our results reveal that the suppressed thermal transport behavior in these materials stems from exceptionally strong three- and four-phonon anharmonic scattering, driven by abundant hierarchical rotation modes. Remarkably, four-phonon scattering rates even surpass those of three-phonon processes for numerous modes, which significantly suppresses thermal transport. Such behavior originates from the large four-phonon scattering phase space induced by phonon quasi-flat bands and wide bandgaps as well as the strong quartic anharmonicity accompanied by rotation modes. Moreover, coherent thermal conductivity contributes significantly in AgB(CN)$_{4}$ and NaB(CN)$_{4}$, with its contribution exceeding that of the particle-like term at elevated temperatures. This leads to a positive temperature dependence of thermal conductivity in NaB(CN)$_{4}$ at high temperature regions. However, certain extreme phonon modes of NaB(CN)$_{4}$ enter the overdamped regime at high temperatures, which render the standard phonon gas model invalid. Therefore, we further employ homogeneous nonequilibrium molecular dynamics (HNEMD) simulations to provide a more robust description for thermal transport. The HNEMD simulations confirm the unified transport theory results at low temperatures but reveal a continuing decrease in $\kappa_{\mathrm{L}}$ at high temperatures, underscoring the limitations of the standard phonon gas model in giant anharmonic systems. Finally, we extend HNEMD simulations to 15 CFMs, including Zn(CN)$_{2}$, Cd(CN)$_{2}$, NaB(CN)$_{4}$, LiX(CN)$_{4}$, CuX(CN)$_{4}$, and AgX(CN)$_{4}$ (X=B, Al, Ga, In). As the X atom transitions from B to In, the corresponding system exhibits weaker bonding and greater root-mean-square displacement, thereby inducing stronger anharmonic scattering and lower $\kappa_{\mathrm{L}}$. Among them, Cd(CN)$_{2}$, NaB(CN)$_{4}$, LiIn(CN)$_{4}$, and AgX(CN)$_{4}$ exhibit ultralow room-temperature $\kappa_{\mathrm{L}}$ values ranging from 0.35 to 0.81 W/mK. These results shed light on the physical mechanisms of thermal transport in CFMs, and suggest a practicable strategy for developing novel lightweight materials with ultralow thermal conductivity.

\section{Methods}\label{sec:4}
All density-functional theory (DFT) calculations are implemented using the Vienna Ab initio Simulation Package (VASP) \cite{RN65} with the projector augmented wave (PAW) \cite{RN66} method. The PBE \cite{RN67} exchange-correlation functional is used to relax the geometric structure. The kinetic energy cutoff for wave function is set to 500 eV, and the k-mesh of 8 × 8 × 8 is used to sample the Brillouin Zone. The DFT-D3 correction \cite{RN68} is adopted to account for van der Waals (vdW) interactions. The ALAMODE package is used to compute both the harmonic second-order interatomic force constants (IFCs) via finite displacement method and the anharmonic higher-order IFCs through compressive sensing lattice dynamics \cite{RN36,RN50,RN69}. High precision multi-atomic cluster expansion (MACE) machine learning potential is constructed to capture IFCs more efficiently \cite{RN70}. Self-consistent phonon (SCP) \cite{RN36,RN50} calculation is performed to obtain the temperature-dependent phonon dispersion and corresponding second-order IFCs. The lattice thermal conductivity is calculated by a unified thermal transport theory and implemented in the modified FOURPHONON software \cite{RN37,RN71,RN72,RN73}. The equation for $\kappa_{\mathrm{L}}$ is as follows
\begin{equation}
	\begin{split}
		\kappa_{\mathrm{L}}^{\mathrm{P/C}} = \frac{\hbar^{2}}{k_{\mathrm{B}}T^{2}VN} \sum_{\mathbf{q}} \sum_{j,j^{\prime}} \frac{\omega_{\mathbf{q}j}+\omega_{\mathbf{q}j^{\prime}}}{2} \nu_{\mathbf{q}jj^{\prime}} \otimes \nu_{\mathbf{q}jj^{\prime}} \\
		\times \frac{\omega_{\mathbf{q}j}n_{\mathbf{q}j}(n_{\mathbf{q}j}+1) + \omega_{\mathbf{q}j^{\prime}}n_{\mathbf{q}j^{\prime}}\left(n_{\mathbf{q}j^{\prime}}+1\right)}{4\left(\omega_{\mathbf{q}j} - \omega_{\mathbf{q}j^{\prime}}\right)^{2} + \left(\Gamma_{\mathbf{q}j} + \Gamma_{\mathbf{q}j^{\prime}}\right)^{2}} \left(\Gamma_{\mathbf{q}j} + \Gamma_{\mathbf{q}j^{\prime}}\right),
	\end{split}
	\label{eq:eq-1}
\end{equation}
where the superscripts P and C represent the contributions from populations (particle-like propagation) and coherences (wave-like phonon tunneling), respectively. When $j$$=$$j^{\prime}$, Eq.~\ref{eq:eq-1} is equivalent to the Peierls-Boltzmann transport equation, $i.e$., the particle-like transport term. Otherwise, it corresponds to the wave-like transport term. Moreover, the $\kappa_{\mathrm{L}}$ is also obtained using homogeneous nonequilibrium molecular dynamics (HNEMD) simulations based on the trained neuroevolution potential (NEP) and implemented in the GPUMD software \cite{RN45,RN74,RN75}. To eliminate the size effects, the large supercell exceeding 85 × 85 × 85 Å$^{3}$ (containing 40960 atoms) is adopted. Corresponding supercell is also applied to simulate the thermal expansion behavior. More computational details are provided in the Supporting Information (SI).

\section*{ASSOCIATED CONTENT}
\noindent{\textbf{Supporting Information}} The Supporting Information is available free of charge at ***

Detailed method description, lattice constant, root-mean-square error of machine learning potential models, training parameters for fitting machine learning potential models, thermal expansion, phonon spectrum, rotation mode eigenvectors, position markers of rotation modes within the phonon spectrum, weighted scattering phase space, and phonon lifetime.

\section*{Data availability}
The data supporting the findings of this study are available within this article and its Supporting Information. The trained NEP and MACE models are openly available in the repository \url{https://github.com/ZhunyunTang/CFMs-ML}. Additional data are available from the corresponding author on reasonable request.

\section*{Code availability}
The source code for GPUMD is available at \url{https://github.com/brucefan1983/GPUMD}, for ALAMODE is available at \url{https://github.com/ttadano/alamode}, for SHENGBTE is available at \url{https://www.shengbte.org}, and FOURPHONON is available at \url{https://github.com/FourPhonon/FourPhonon}, for PYNEP is available at \url{https://github.com/bigd4/PyNEP}, for the modified version of FOURPHONON is available at \url{https://github.com/ZhunyunTang/FourPhonon--wigner-sampling}.

\section*{ACKNOWLEDGMENTS}

This work was supported by the National Natural Science Foundation of China (Grant No. 52372260), the Science Fund for Distinguished Young Scholars of Hunan Province of China (No. 2024JJ2048), the Hunan Provincial Innovation Foundation for Postgraduate (No. CX20240058), and the Youth Science and Technology Talent Project of Hunan Province (No. 2022RC1197). The calculation in this work is partly supported by the high-performance computing platform of school of physics and optoelectronics.

\section*{Author contributions}
Z.T. contributed to the methodology, software, all calculations, formal analysis, visualization, and writing of original draft.  X.W., J.L., C.H., and M.C. contributed to discussions and writing of manuscript. C.T. and T.O. were responsible for supervision, writing and discussion.

\section*{Competing interests}
The authors declare no competing interests.

\bibliography{reference}
\end{document}